\documentstyle[12pt,epsf]{article}

\title{Quantum thermodynamic fluctuations of a chaotic Fermi--gas model}

\author{P. Leboeuf$~ ^1$ \ and A. G. Monastra$^2$}

\newcommand{\tauH}{\tau_{\scriptscriptstyle H}}

\newcommand{\geqa}{\stackrel{>}{\scriptstyle \sim}}
\newcommand{\tmin}{\tau_{\rm min}}
\newcommand{\Tc}{T_{\rm c}}
\newcommand{\Ec}{E_{\rm c}}

\begin{document}
\maketitle

\begin{center}
\small{1. \ Laboratoire de Physique Th\'eorique et Mod\`eles
Statistiques,\footnote[3]{Unit\'e de Recherche de l'Universit\'e Paris XI
associ\'ee au CNRS} \\ B\^at. 100, Universit\'e de Paris-Sud, 91405
Orsay Cedex, France}
\end{center}

\begin{center}
\small{2. \ Department of Physics of Complex Systems, The Weizmann
Institute of Science,\\ 76100 Rehovot, Israel}
\end{center}

\baselineskip 0.25in

\hspace{1.5in}

\begin{abstract}
We investigate the thermodynamics of a Fermi gas whose single--particle energy
levels are given by the complex zeros of the Riemann zeta function. This is a
model for a gas, and in particular for an atomic nucleus, with an underlying
fully chaotic classical dynamics. The probability distributions of the quantum
fluctuations of the grand potential and entropy of the gas are computed as a
function of temperature and compared, with good agreement, with general
predictions obtained from random matrix theory and periodic orbit theory
(based on prime numbers). In each case the universal and non--universal
regimes are identified.
\end{abstract}

\vspace{2cm}


\pagebreak

\section{Introduction} 
\label{sec:Intro}
\setcounter{equation}{0}

Fermi gases confined to small volumes present distinct finite size effects in
their thermodynamic properties. Among the different corrections to the bulk
behavior, some are smooth functions of the parameters of the system. They
typically depend on geometric aspects of the volume occupied by the gas (e.g.
surface and curvature terms). In contrast, other corrections are fluctuating
functions. These quantum (or mesoscopic) fluctuations are related to the
discreteness of the single--particle spectrum, and are sensitive to the nature
of the dynamics of the particles in the gas. In many cases they are small
compared to the bulk, but may play nevertheless an important role (like, for
example, in nuclear physics where shell effects are crucial in the
determination of the shape of atomic nuclei \cite{bm,sm}). In other (more
spectacular) situations, they provide the dominant contribution. This occurs
for example in the electronic orbital magnetism in quantum dots -- where they
can be larger than the Landau susceptibility \cite{shapiro,voppen,ruj} --, or
in the persistent currents in mesoscopic rings \cite{agi,sch} -- where the
bulk contribution vanishes. In all cases the quantum fluctuations disappear
when the temperature is raised.

Instead of a detailed computation and description of the quantum fluctuations
for a particular system, the aim here is to study their statistical
properties. The interest of such an analysis is well known: using a minimum
amount of information, a statistical approach allows to establish a
classification scheme among the fluctuations of different physical systems. It
also allows to distinguish the generic from the specific, and provides a
powerful predictive tool in complex systems. A general description of the
statistical properties of the quantum fluctuations of thermodynamic functions
of integrable and chaotic ballistic Fermi gases, and of their temperature
dependence, was developed in Ref.\cite{lm3}.

From a classical point of view, there are two extreme cases of
single--particle motion, namely integrable and fully chaotic. The case of a
mixed phase space (e.g., coexistence of regular and chaotic motion) being the
generic one. In all cases, in a Fermi gas model the quantum thermodynamic
fluctuations can be directly related to the fluctuations of the
single--particle, discrete, energy levels. Schematically, there are two
distinct types of single--particle fluctuations. On the one hand there are the
local fluctuations, i.e. those occurring on scales of the order of the
single--particle mean level spacing $\delta$. For the two extreme dynamics
mentioned above, these fluctuations are known to be universal \cite{bg},
namely Poisson for integrable motion, random matrix in full chaos. The second
type of fluctuations generically present in any single--particle spectrum are
long range correlations, that occur on a scale $E_c$, where $\Ec$ is the
energy associated with the shortest classical periodic orbit of the mean field
(inverse time of flight across the system). In contrast to the local
fluctuations, the long range modulations depend on the short--time specific
properties of the dynamics and are therefore non--universal, either for
integrable, mixed, or fully chaotic dynamics (thought their importance varies
in each case).

Properties related to both types of single--particle fluctuations have been
identified in the nuclear behavior. For example, shell effects in the nuclear
masses are a clear manifestation of the long range correlations, whereas
nearest--neighbor spacing statistics and the distribution of widths of neutron
resonances illustrate universal local fluctuations. However, a global and
comprehensive picture of the quantum fluctuations has not yet emerged. An
important additional ingredient are certainly the variations in the nucleus of
the nature of the dynamics as a function of the excitation energy. But even in
the case of a scaling mean--field motion where the classical dynamics is
energy independent (like when putting a Fermi gas inside a hard wall box or
billiard) the physical behavior of the gas can be quite complex and rich, and
both types of fluctuations (local and long range) can in fact manifest in
different quantities related to the system, or in the same quantity at
different temperatures \cite{lm3}.

One might think that a fully chaotic single--particle dynamics implies
universal quantum fluctuations of the thermodynamics of the gas, well
described by some random matrix model (Wigner, two--body random, etc). This
belief is not true in general. The reason is that the thermodynamic
fluctuations are universal only when they are controlled by the local
single--particle fluctuations. And this is not always the case. For example,
it has been shown, in particular, that the statistical properties of the
ground--state energy of a many--body system are controlled by the short
periodic orbits (at least in systems with a ground state well described by a
mean--field approximation, like the atomic nucleus) \cite{lm3}. The
corresponding distribution is therefore non--universal. As a consequence, it
makes no sense to model the ground--state fluctuations by some random matrix
model, that possess no information on the specific mean--field short time
properties of the dynamics.

Our purpose here is to illustrate the richness of the thermodynamic quantum
fluctuations and to explicitly check some of the predictions by considering a
particular example of a Fermi gas with a fully chaotic classical dynamics. The
system considered is a mathematical model. It is achieved by considering a
spectrum of single--particle energy levels constructed from the imaginary part
of the complex zeros of the Riemann zeta function. The Fermi gas is obtained
by filling the single--particle energy levels with an average occupation
number determined by the Fermi--Dirac distribution. This model of a chaotic
gas was introduced in Ref.\cite{blm1}, where general motivations as well as
dynamical analogies were given. If this Fermi--gas model is viewed as an
element of the periodic table, following nuclear physics terminology we
referred to it as the {\it Riemannium} \cite{blm1}. Aside its mathematical
interest, physically the Riemannium is an important model because it possesses
all the typical dynamical properties and characteristics of a realistic
chaotic gas, while it greatly facilitates the numerical and analytical
computations (e.g., all the ingredients required in a semiclassical
description of its thermodynamic properties are known). This is quite unusual
in fully chaotic systems, and therefore offers a rather unique model to test
general results and ideas.

The basic quantities to be computed are the probability distributions, at a
given temperature, of the quantum fluctuations of the grand potential and of
the entropy of the gas. We consider a degenerate gas in the grand canonical
ensemble; other ensembles (e.g., canonical) may be considered as well. In
particular, it can be shown that to leading order in an expansion of large
chemical potential, the fluctuations of the energy of the gas at a fixed
number of particles coincide with those of the grand potential. Our results
therefore apply to the total energy. In a semiclassical approximation, the
fluctuations are described by the periodic orbits of the classical dynamics
associated with the single--particle motion \cite{bb,gutz,sm,brack}. In the
Riemannium, the role of the periodic orbits is played by the prime numbers.
The basic equations and the relevant energy scales of the Riemannium are
introduced in \S \ref{sec:Maineq}. In \S \ref{sec:Omega} we analyze the
fluctuations of the grand potential. The corresponding distribution is
non--universal at all temperatures (including $T=0$, the case considered in
\cite{blm1}). The asymptotic moments of the distribution, obtained for a large
chemical potential, are well reproduced by some convergent sums dominated by
the smaller prime numbers (i.e., short periodic orbits). Odd moments are
non--zero, implying an asymmetric distribution at all temperatures. At a
finite chemical potential, large prime numbers introduce corrections to the
asymptotic moments. These corrections to the non--universal leading--order
behavior are, in contrast, universal, and described by random matrix theory.

The entropy fluctuations are studied in \S \ref{sec:Entropy}. Their character
strongly depends on temperature. For a chaotic dynamics without time reversal
symmetry, which corresponds to the Riemannium, these fluctuations present
several remarkable features, predicted in \cite{lm3}. For temperatures $T \ll
\delta$, the variance of the entropy fluctuations increases linearly. In a
second regime, when $\delta \ll T \ll \Ec$, the variance saturates to a
constant value. This initial behavior of the variance (i.e. linear growth
followed by a saturation), and more generally of the probability distribution
of the entropy fluctuations, was predicted to be universal, namely identical
for any chaotic system without time reversal symmetry (with a given
single--particle mean level spacing at Fermi energy). It is well described by
the fluctuations obtained from a GUE random matrix single--particle spectrum.
The shape of the probability distribution strongly depends on temperature, and
converges towards a Gaussian as temperature increases. However, this universal
behavior disappears for temperatures of the order $\Ec$ or higher, where the
statistical properties of the gas do not coincide any more with random matrix
theory. The probability distribution of the fluctuations is now described by
the short periodic orbits, which are system dependent. In this high
temperature regime the typical size of the fluctuations shows an overall
exponential decay. All these predictions are worked out explicitly for the
Riemannium and compared with numerical simulations. We have found very good
agreement between theory and numerical experiments in all the regimes.

Although only two particular thermodynamic functions are discussed, they
illustrate all the rich possibilities that may be encountered in other
functions. Similar methods apply to them.

%%%%%%%%%%%%%%%%%%%%%%%%%%%%%%%%%%%%%%%%%%%%%%%%%%%%%%%%%%%%%%%%%%%

\section{Thermodynamic framework}
\label{sec:Maineq}
\setcounter{equation}{0}

In the grand canonical ensemble the equilibrium properties of a Fermi gas are
defined by the grand potential
\begin{equation}
\Omega (\mu,T) = - T \int dE ~ \log [1+e^{(\mu-E)/T}] ~ \rho(E) \ ,
\label{omegint}
\end{equation}
where $T$ is the temperature (we set Boltzmann's constant $k_B =1$), $\mu$ the
chemical potential, and
\begin{equation}
\rho (E) = \sum_j \delta (E - E_j) 
\end{equation}
is the single--particle density of states. In the Riemannium, the
single--particle energies are given by the imaginary part of the upper--half
complex zeros $s=1/2 + i E_j$ of the Riemann zeta function $\zeta (s)$, $ E_j
= 14.1347, 21.0220, 25.0109, \ldots$ (the values used in all our numerical
simulations were obtained from \cite{odlyzko}). They are expressed in
arbitrary units, which are also those of temperature (since $k_B =1$). From
the grand potential, other thermodynamic functions can be computed. We are
particularly interested in the behavior of the entropy of the gas,
\begin{equation} \label{Sdef}
S(\mu,T) = -\left( \frac{\partial \Omega}{\partial T} \right)_{\mu} \ . 
\end{equation}

The number of complex zeros with imaginary part less than $E$ is
\cite{edwards}, 
\begin{equation} 
n(E) = 1 - \frac{1}{2\pi} E \log \pi - \frac{1}{\pi} Im \log \Gamma \left(
\frac{1}{4} - i \frac{E}{2} \right) - \frac{1}{\pi} Im \log \zeta \left(
\frac{1}{2} - i ~ E \right) \ . \label{nzexact}
\end{equation}
Expanding the $\Gamma$ function and using the Euler product representation of
$\zeta$ in the last term, this expression splits naturally into a smooth plus
an oscillatory part, as in the semiclassical approximation in quantum systems.
The density of zeros $\rho (E) = d n / d E$ is written $ \rho (E) =
\overline{\rho} (E) + \widetilde{\rho} (E) $, with
\begin{eqnarray} \label{rhobar}
\overline{\rho} (E) &=& \frac{1}{2\pi} \log \left( \frac{E}{2\pi}
\right) + {\cal O} (E^{-2}) \ , \\
\widetilde{\rho} (E) &=& - \frac{1}{\pi} \sum_p \sum_{r=1}^{\infty}
\frac{\log p}{p^{r/2}} \cos( E ~ r ~ \log p) \ . \label{rhoosc}
\end{eqnarray}
In the Riemannium, the first term corresponds therefore to the asymptotic
average density of the single--particle energy levels. In the fluctuating part
$\widetilde{\rho}$, the sum is made over the prime numbers $p$. The comparison
of this equation with the semiclassical Gutzwiller trace formula for the
spectral density of a dynamical system \cite{gutz}, $\widetilde{ \rho } (E) =
\sum_{{\rm p.o.}} A_{{\rm p.o.}} \cos (S_{{\rm p.o.}}(E)/\hbar)$ (the sum is
over the classical periodic orbits), shows that each prime number may be
interpreted as the label of an unstable primitive periodic orbit $p$ of a
fully chaotic system without time reversal symmetry, of action $S_p = E \,
\tau_p$, period $\tau_p = \log p$, Lyapounov stability $\lambda_p = 1$,
repetitions labeled by $r$, and $\hbar=1$ \cite{berry2,foot}. Because of this
mapping, in the following we often refer to the prime numbers as periodic
orbits. Putting aside the issue of the very existence of a classical system
whose quantum eigenvalues coincide with the imaginary part of the Riemann
zeros, it is the formal analogy with the semiclassical theory of dynamical
systems that makes the Riemann zeros in general, and the Riemannium in
particular, interesting on a physical basis as a model to test general
theories of quantum chaotic systems. Additional support for a dynamical
interpretation comes from numerical simulations, that clearly indicate
\cite{odlyzko} that the sequence of the imaginary part of the Riemann zeros
obeys the Bohigas--Giannoni--Schmit conjecture for chaotic systems without
time--reversal symmetry \cite{bgs,bg}: asymptotically, their statistics
coincide with those of eigenvalues of the GUE ensemble of random matrices
\cite{mont}.

Replacing the decomposition (\ref{rhobar})--(\ref{rhoosc}) of the density of
states into (\ref{omegint}) leads to a corresponding decomposition of the
grand potential. A similar decomposition for other thermodynamic functions is
obtained by derivation. We are interested in the statistical properties of the
fluctuating part of those functions, and in their temperature dependence. In
the forthcoming analysis, the different relevant scales are the following.
\begin{itemize}
\item[i)] Largest scale: chemical potential $\mu$
\item[ii)] Intermediate scale: shortest periodic orbit
\begin{eqnarray}\label{ec}
&~& \hspace*{-2cm} {\rm Energy:} \ \Ec = h / \tmin = 2 \pi / \log 2
\approx 9.06472 \ , \nonumber \\ &~& \hspace*{-2cm} {\rm Time:} \ \tmin =
\log 2 \ , \\ &~& \hspace*{-2cm} {\rm Temperature:} \ \Tc = \Ec / 2 \pi^2 
\approx 0.459224 \nonumber \ .
\end{eqnarray}
\item[iii)] Smallest scale: single--particle mean level spacing
\begin{eqnarray}\label{delta}
&~& \hspace*{-2cm} {\rm Energy:} \ \delta = 1 / \overline{\rho} = 2 \pi / \log 
(\mu/2\pi) \ , \nonumber \\
&~& \hspace*{-2cm} {\rm Time:} \ \tauH = h/\delta = \log (\mu/ 2\pi)  \ , \\
&~& \hspace*{-2cm} {\rm Temperature:} \ T_{\delta} = \delta / 2 \pi^2 = 
1 / \pi \log (\mu/2\pi) \nonumber \ .
\end{eqnarray}
\end{itemize}
The intermediate scale is related to the shortest periodic orbit (or time of
flight across the system at energy $\mu$) which, according to the previous
analogy, corresponds here to the period $\tmin = \log 2$ associated with the
prime number 2. $\Ec$ defines the outer energy scale of oscillations in the
density of states (cf Eq.(\ref{rhoosc})). Due to the independence of the
periods with energy in the Riemann dynamics, this energy scale is a fixed
number. In the definition of the mean level spacing $\delta$, we have used the
asymptotic leading order approximation of the average density of states.
$\tauH$ is the so--called Heisenberg time. The $2 \pi^2$ factor (approximately
20) that relates energies and temperatures is dictated by semiclassical
arguments \cite{lm3}, and it has to be taken into account in numerical
comparisons. A final important parameter is the ratio of $\Ec$ to $\delta$,
which measures the number of single--particle states contained in a scale
$\Ec$,
$$
g=\Ec /\delta = \log (\mu/2\pi)/\log 2 \ .
$$ 
In the semiclassical limit $\mu \rightarrow \infty$, the three energy scales
are well separated, $ \delta \ll \Ec \ll \mu$, and $g \rightarrow \infty$.

%%%%%%%%%%%%%%%%%%%%%%%%%%%%%%%%%%%%%%%%%%%%%%%%%%%%%%%%%%%%%%%%%%%

\section{The grand potential}
\label{sec:Omega}
\setcounter{equation}{0}

Replacing $\rho (E)$ by $\overline{\rho}$ in Eq.(\ref{omegint}), integrating
by parts, and applying Sommerfeld's approximation (valid when $T \ll \mu$), we
obtain the usual thermodynamic relation
\begin{equation}
\overline{\Omega} ( \mu , T ) = \overline{\Omega}_0 (\mu) - 
\frac{\pi^2}{6} ~ \overline{\rho} (\mu) T^2 \ , \label{omegliso}
\end{equation}
where $\overline{\Omega}_0$ is the grand potential at zero temperature. For
the Riemannium
$$
\overline{\Omega}_0 (E) = -\frac{E^2}{4 \pi} \log \left( \frac{E}{2\pi} \right)
+ \frac{3}{8 \pi} E^2 - \frac{7}{8} E - \frac{\log E}{48 \pi} + c + {\cal O}
(E^{-2}) \ .
$$
Terms of order up to $E^{-2}$ in $\bar{\rho}$ have been included in the
integration. The determination of the constant $c$ is of numerical importance
when computing the fluctuations. From the work of Selberg on the function
$S_1(t)$ \cite{selberg} (which, up to an overall sign, coincides with the
fluctuating part of the grand potential at zero temperature), we can extract
its value
$$
c = \frac{1}{\pi} \int_{1/2}^{\infty} \log \left| \zeta (x) \right| dx
- \lim_{E \rightarrow \infty} \left\{ \int_{0}^{E} \overline{n} (x) \ dx
- \frac{E^2}{4 \pi} \log \left( \frac{E}{2\pi} \right) + \frac{3}{8
\pi} E^2 - \frac{7}{8} E - \frac{\log E}{48 \pi} \right\} ,
$$
where $\overline{n} (x)$ is the exact smooth part of the counting function,
computed from equation (\ref{nzexact}). Evaluating numerically the integrals,
the value of the constant is found to be $c \approx 0.75575$.

Replacing $\rho (E)$ by (\ref{rhoosc}) in Eq.(\ref{omegint}), the oscillating
part of the grand potential is
\begin{equation}
\widetilde{\Omega} ( \mu , T ) = - \frac{1}{\pi} \sum_p
\sum_{r=1}^{\infty} \frac{\kappa ( \pi \ T \ r \log p) }{ r^2 \ p^{r/2}
\log p} \cos( \mu \ r \log p) \ . \\ \label{Oosc}
\end{equation}
The function 
$$ 
\kappa (x) = x / \sinh x 
$$ 
takes into account temperature effects \cite{sm}, and cuts exponentially the
contribution of large prime numbers. If $T > \Tc$, the contribution of each
orbit, including the shortest one, is exponentially small.

Equation (\ref{Oosc}) is an oscillatory function of $\mu$ that describes the
quantum fluctuations of the grand potential. To study their statistical
properties we define the average
\begin{equation}
\langle f \rangle \equiv \frac{1}{\Delta \mu} \int_{\mu - \Delta \mu /2}^{
\mu + \Delta \mu / 2 } f(\mu') ~ d \mu' \ , \label{mean}
\end{equation}
for any thermodynamic function $f(\mu)$ associated to the Riemannium. The size
of the average window $\Delta \mu$ has to be much smaller than $\mu$ (in order
to extract local properties), but large enough to contain a sufficient number
of oscillations to make the statistical approach meaningful, i.e. $ \Ec \ll
\Delta \mu \ll \mu $.

%%%%%%%%%%%%%%%%%%%%%%%%%%%%%%%%%%%%%%

\subsection{Variance}\label{secvaro}

The most basic feature of the probability distribution of the fluctuations is
the variance. The average of the square of $\widetilde{\Omega}$ can be written
as an integral \cite{lm3}
\begin{equation}
\langle \widetilde{\Omega}^2 \rangle (\mu , T ) = \frac{1}{2 \pi^2}
\int_{0}^{\infty} d\tau ~ \frac{ \kappa^2 ( \pi T \tau ) }{ \tau^4 } ~
K (\tau) \ . \label{varint}
\end{equation}
The function
\begin{equation} \label{ff}
K (\tau) = \left\langle \sum_{i,j} A_i A_j \cos[ \mu (r_i \log p_i - r_j
\log p_j) ] \delta ( \tau - \bar{\tau} ) \right\rangle \ ,
\end{equation}
is the form factor, i.e. the Fourier transform of the two--point correlation
function of the single--particle spectrum; $A_i = \log p_i/ p_i^{r_i / 2} $,
and $ \bar{\tau} = ( \tau_{p_i}+ \tau_{p_j} ) / 2 $. This function is central
in our analysis. In the semiclassical limit $g \rightarrow \infty$ it is
conjectured to coincide with the GUE form factor
\begin{equation}\label{kgue}
K_{GUE} (\tau) = {\rm min} \{ \tau, \tauH \} \ . 
\end{equation}
If we naively replace in (\ref{varint}) $K (\tau)$ by $K_{GUE} (\tau)$, the
integral diverges at $\tau \rightarrow 0$. This is because at finite $g$
random matrix theory provides a wrong description of the short--time behavior
of the form factor. For low values of $\tau$ the diagonal terms in
Eq.(\ref{ff}) should be used \cite{berry1},
\begin{equation}\label{kd}
K_D (\tau) = \sum_{i} A_i^2 \ \delta ( \tau - \tau_i )  \ . 
\end{equation}
This shows that the form factor is non--universal at short times, and that
$K(\tau) = 0$ for $\tau < \tmin$. The form factor behaves like $K_{GUE}$ only
for times larger than some $\tau_*$. The intermediate time $ \tau_* $
satisfies $({\rm Re} \gamma)^{-1} < \tau_* < \tauH $, where $\gamma$ is the
classical resonance (eigenvalue of the Perron--Frobenius operator) with the
smallest real part \cite{agam}.

Since short times dominate the fluctuations of the grand potential, in the
limit $\tau \rightarrow 0$ we use in (\ref{varint}) the diagonal approximation
of the form factor, and neglect for the moment the contributions of longer
times,
\begin{equation}
\langle \widetilde{\Omega}^2 \rangle_{{\rm D}} = \frac{1}{2 \pi^2}
\sum_{ p , r} \frac{ \kappa^2 ( \pi T \ r \log p ) }{ r^4 ~ p^r ~
\log^2 p } \ . \label{varodiag}
\end{equation}
This is a convergent sum for all temperatures, and is independent of the
chemical potential. At $T = 0$ it gives the value $\langle
\widetilde{\Omega}^2 \rangle_{{\rm D}} \approx 0.079290$ \cite{blm1}. Since
the temperature factor is a monotonically decreasing function of $\tau$, the
variance exponentially decreases with increasing temperature, and the role of
the short orbits (small primes) in the fluctuations is further enhanced. Fig.1
compares Eq.(\ref{varodiag}) evaluated at different temperatures to numerical
results obtained for the Riemannium (for reference, in the chemical potential
window used to compute the fluctuations we have $T_\delta \approx 0.013$).

The agreement with the leading order description of the variance is excellent.
However, as the chemical potential is lowered deviations from the asymptotic
behavior are observed. The corrections may be obtained from Eq.(\ref{varint}),
$\langle \widetilde{\Omega}^2 \rangle = \langle \widetilde{\Omega}^2
\rangle_{{\rm D}} + \langle \widetilde{\Omega}^2 \rangle_{{\rm off}}$, by
using the GUE form factor (\ref{kgue}) for times $\tau > \tau_*$. This gives,
\begin{equation}
\langle \widetilde{\Omega}^2 \rangle_{{\rm off}} = \frac{1}{2 \pi^2}
\int_{\tauH}^{\infty} d\tau ~ \frac{ (\tauH - \tau ) ~ \kappa^2 ( \pi
T \tau ) }{ \tau^4 } \ .
\end{equation}
As temperature goes to zero, the temperature factor $\kappa^2$ tends to 1, and
the integral is easily performed giving $ \langle \widetilde{\Omega}^2
\rangle_{{\rm off}} = - 1 / (12 \pi^2 \tauH^2) $. This provides a
finite--$\mu$ correction to the variance, which decreases as $(\log
\mu)^{-2}$. For increasing temperatures the correction is smaller. The
off-diagonal terms thus provide a universal correction to the leading non
universal diagonal term (namely, the only specific information on the system
that enters the correction is the average density of states at Fermi energy,
through $\tauH = h \overline{\rho}$).

%%%%%%%%%%%%%%%%%%%%%%%%%%%%%%%%%%%

\subsection{Higher moments and distribution}
\label{sec:moments}

We already showed that to calculate the variance the diagonal approximation is
accurate because the short orbits dominate the fluctuations. A generalization
of the diagonal approximation allows to evaluate all the moments of the
probability distribution of the grand potential by a method developed in
\cite{lm3} and \cite{blm1}. The mechanism is an interference process between
repetitions of primitive periodic orbits.

Defining the amplitude
\begin{equation}\label{apr}
{\cal A}_{p,r} (T) = - \frac{1}{2 \pi} \frac{\kappa ( \pi \ T \ r \log
p) }{ r^2 \ p^{r/2} \log p} \ ,
\end{equation}
the third and fourth moments of the distribution of $\widetilde{\Omega}$ are
found to be, in the diagonal approximation,
\begin{eqnarray}
\langle \widetilde{\Omega}^3 \rangle &=& 6 ~ \sum_p \sum_{r_1, r_2 =
1}^{\infty} {\cal A}_{p,r_1} ~ {\cal A}_{p,r_2} ~ {\cal A}_{p , r_1 +
r_2} \ , \label{m3} \\
\langle \widetilde{\Omega}^4 \rangle &=& 2 ~ \sum_p \Big[ ~ 4
\sum_{r_1, r_2, r_3 = 1}^{\infty} {\cal A}_{p, r_1} ~ {\cal A}_{p,
r_2} ~ {\cal A}_{p, r_3} ~ {\cal A}_{p, r_1 + r_2 + r_3} - ~ 6
\sum_{r_1, r_2=1}^{\infty} {\cal A}^2_{p, r_1} ~ {\cal A}^2_{p, r_2}
\nonumber \\
& ~ & + ~ 3 \sum_{r_1, r_2 = 1}^{\infty} \sum_{r_3 = 1}^{r_1 + r_2 -
1} {\cal A}_{p, r_1} ~ {\cal A}_{p, r_2} ~ {\cal A}_{p, r_3} ~ {\cal
A}_{p, r_1 + r_2 - r_3} \Big] + 3 ~ \langle \widetilde{\Omega}^2
\rangle^2 . \label{m4}
\end{eqnarray}
All the sums are convergent. We see explicitly that the third moment and the
excess of the fourth moment (the term $3 ~ \langle \widetilde{\Omega}^2
\rangle^2$ being the fourth moment of a Gaussian distribution) are different
from zero. The distribution of the fluctuations of the grand potential is
therefore a non universal asymmetric function that strongly depends on the
small prime numbers. It is displayed, for different temperatures, in Fig.2
\cite{foot2}.

Higher moments $k \geq 5$ may also be computed, with increasing complexity of
the result. Corrections to the diagonal terms coming from off--diagonal
contributions exist, but as already shown they are negligible in the limit
$\mu \rightarrow \infty$. We thus ignore them.

These results are confirmed to high accuracy by the numerical data from the
Riemann zeros (cf Fig.1 and Table 1).

%%%%%%%%%%%%%%%%%%%%%%%%%%%%%%%%%%%%%%%%%%%%%%%%%%%%%%%%%%%%%%%%%%%

\section{The entropy}
\label{sec:Entropy}
\setcounter{equation}{0}

From Eq.(\ref{Sdef}) the entropy of the gas is expressed as,
\begin{equation}
S(\mu,T) = \int s \left( \frac{\mu - E}{T} \right) \rho (E) ~ dE =
\sum_j s \left( \frac{\mu - E_j}{T} \right) \ ,
\label{Sint}
\end{equation}
where
\begin{equation}
s (x) = \log ( 1 + e^x ) - \frac{x}{1+e^{-x}} \ .
\end{equation}
The function $s(x)$ is localized around $x=0$, and decays exponentially for
$|x| \gg 1$. At very low temperatures ($T \ll T_\delta$), the function
$S(\mu,T)$ has peaks as a function of $\mu$ centered at each single particle
energy level (Riemann's zero) of width $\ll \delta$. At these temperatures the
entropy is therefore zero for $\mu \neq E_j$, and takes the value $\log 2$
when $\mu = E_j$. As the temperature increases, the width increases (while the
height remains constant) and the peaks start to overlap. At temperatures $T
\sim T_\delta$ and at a fixed $\mu$, only the energy levels distant by a few
mean spacings from the chemical potential contribute to the entropy. The
fluctuations are governed by the local statistics of the eigenvalues, which
are universal (e.g., GUE). In this regime universality in the entropy
fluctuations is expected. In contrast, at higher temperatures $T \geqa \Tc$,
when peaks separated from $\mu$ by a distance of order $\Ec$ start to
contribute to the entropy, the universality will be lost because information
on the scale of the shortest periodic orbit enters. We now substantiate this
analysis by an explicit quantitative calculation.

Deriving with respect to temperature Eqs. (\ref{omegliso}) and (\ref{Oosc}),
the smooth and fluctuating part of the entropy are given, respectively, by
\begin{eqnarray}
\overline{S} ( \mu , T ) &=& \frac{\pi^2}{3} \ \overline{\rho} ( \mu )
\ T \ , \\ \label{Slis}
\widetilde{S} ( \mu , T ) &=& \sum_p \sum_{r=1}^{\infty} \frac{
\kappa' ( \pi \ T \ r \log p) }{ r \ p^{r/2} } \cos( \mu \ r \log p) \
. \label{Sosc}
\end{eqnarray}
The function $\kappa' (x) = d \kappa / d x $ decreases also exponentially for
$x \gg 1$, producing a cut-off for prime numbers satisfying $\log p \gg 1 / (
\pi T)$. When $x \rightarrow 0$, $\kappa' (x)$ vanishes linearly. The maximum
of the function is located at $x \approx 1.6061$. Prime numbers that satisfy
$r \log p \sim 1 / ( \pi T)$ give therefore the main contribution to the
oscillations. At low temperatures large primes (long orbits) are selected.
Smaller primes (e.g., short orbits) contribute as the temperature is raised.

%%%%%%%%%%%%%%%%%%%%%%%%%%%%%%%%%%%

\subsection{Variance}

Analogously to the grand potential, the variance of the entropy fluctuations
may also be written in terms of the form factor,
\begin{equation}
\langle \widetilde{S}^2 \rangle = \frac{1}{2} \int_{0}^{\infty} d\tau
~ \frac{ \kappa'^2 ( \pi \ T \ \tau ) }{ \tau^2 } ~ K (\tau) \ .
\label{Svar}
\end{equation}
We briefly review here the general results obtained in Ref.\cite{lm3}, and
check them with the Riemannium.

The two different regimes of the form factor described in \S \ref{secvaro}
split the integral (\ref{Svar}) into three different parts,
\begin{equation}
\langle \widetilde{S}^2 \rangle = \frac{1}{2} \left\{ \sum_{r \log p < \tau_*
} \frac{ \kappa'^2 ( \pi T \ r \log p ) }{ r^2 ~ p^r } + \int_{\tau_*}^{\tauH}
d\tau ~ \frac{ \kappa'^2 ( \pi T \tau ) }{ \tau } + \tauH \int_{ \tauH }^{
\infty } d\tau ~ \frac{ \kappa'^2 ( \pi T \tau ) }{ \tau^2 } \right\} \ .
\label{S2}
\end{equation}
The sum corresponds to the non universal short--time regime, whereas
the two integrals correspond to the long--time random matrix behavior.
These different terms dominate the integral at different temperatures.

\vspace{0.2cm}
\noindent
* \underline{Low temperatures}: $T \ll T_\delta$. In this regime the maximum
of $\kappa'$ is centered at times much larger than the Heisenberg time
$\tauH$. The dominant term in Eq.(\ref{S2}) is the last one. We can extend the
integral down to zero with a negligible error. The variance of the entropy is
\begin{equation} \label{s2l}
\langle \widetilde{S}^2 \rangle \approx I_2 \ \overline{\rho} (\mu) \ T =
I_2 \ T/\delta \ ,
\end{equation}
where
\begin{equation}
I_2 = \pi^2 \int_{0}^{\infty} dx ~ \frac{\kappa'^2 ( x ) }{ x^2 }
\approx 1.51836 \ .
\end{equation}
In this initial regime the growth is linear, with a slope proportional to the
density of states. No additional specific information on the Riemann zeta is
present in this formula. Eq.(\ref{s2l}) reflects the discreteness of the
single--particle spectrum, and is thus a very general result valid for
arbitrary systems, independently of their dynamics \cite{lm3}. The smooth part
of the entropy grows also linearly with temperature, and is proportional to
the density of states \cite{landau}. Compared to the mean value
$\overline{S}$, the typical size of the fluctuations $\sqrt{\langle
\widetilde{S}^2 \rangle}$ are large (of order $(T/\delta)^{-1/2})$. In this
regime the quantum fluctuations dominate.

\vspace{0.2cm}
\noindent
* \underline{Intermediate temperatures}: $T_\delta \ll T \ll \Tc$. Now the
maximum of $\kappa'$ is centered at times in between $\tmin$ and $\tauH$.
The dominant term in expression (\ref{S2}) is the second one, where we have
introduced the GUE diagonal approximation of the form factor $K_D (\tau) =
\tau$. Neglecting the contributions of short and long times we extrapolate the
integral from 0 to $\infty$,
\begin{equation} \label{sat}
\langle \widetilde{S}^2 \rangle \approx I_1 = \frac{1}{2}
\int_{0}^{\infty} dx ~ \frac{ \kappa'^2 ( x ) }{ x } \approx
0.0709159 \ .
\end{equation}
In this regime, the size of the entropy fluctuations are therefore insensitive
to temperature variations, they saturate to a universal constant.

\vspace{0.2cm}
\noindent
* \underline{High temperatures}: $T \sim \Tc $ and higher. In this regime the
diagonal approximation of the form factor is still accurate, but the short
time (non universal) structure is now apparent. At this temperatures the
entropy fluctuations are also sensitive to the fact that below $\tmin = \log
2$ the form factor is strictly zero. The dominant term in Eq.(\ref{S2}) is the
first sum. The sum can be extended to all prime numbers and repetitions with
negligible error, and gives a variance
\begin{equation}
\langle \widetilde{S}^2 \rangle \approx \frac{1}{2} \sum_p \sum_{r =
1}^{\infty} \frac{\kappa'^2 ( \pi \ T \ r \log p)}{ r^2 \ p^{r}
} \ .
\end{equation}
For $T \gg \Tc$, all the terms of this sum are exponentially small,
and the fluctuations vanish accordingly.

To have a global description of $\langle \widetilde{S}^2 \rangle$ that
interpolates between the different regimes described above a numerical
evaluation of Eq.(\ref{S2}) is necessary. The result depends on $\mu$ through
the Heisenberg time.

We have checked these predictions by a direct comparison of Eq.(\ref{S2}) to a
numerical computation of the entropy variance as a function of temperature.
The result is displayed in Fig.3 (for reference, $T_\delta \approx 0.013$).
There is an excellent agreement with theory for all temperatures. The initial
linear growth and the saturation to a plateau are amplified in the inset. The
size of the fluctuations almost reaches the theoretical prediction (\ref{sat})
at $T \approx 2.5 \ T_{\delta} \approx 0.03$. The expected intermediate
plateau is however short, because the temperatures $T_\delta$ and $\Tc$ are
not sufficiently well separated (due to the slow logarithmic decrease of
$\delta$, even at these large values of $\mu$ we are not sufficiently
asymptotic). The exponential decay is also well described.

%%%%%%%%%%%%%%%%%%%%%%%%%%%%%%%%%%%

\subsection{Distribution}

In the regime $T \ll \Tc$ the previous results confirm that the behavior of
the statistical properties of the entropy fluctuations are universal. They
depend only on the structure of the GUE form factor and not on any specific
property of Riemann's zeros. The universality is not expected to be valid only
for the variance, but more generally for the full probability distribution. To
check this, we compare the probability distribution of the entropy
fluctuations obtained from Eq.(\ref{Sint}) using two different
single--particle spectra $\{ E_j \}$:
\begin{itemize}
\item[a)] the zeros of the Riemann zeta function,
\item[b)] the eigenvalues of a GUE ensemble of random matrices.
\end{itemize}
In the first case the probability distribution is computed by varying the
chemical potential in a small window, whereas in the second by averaging over
the Gaussian ensemble. Both probability distributions, computed at different
temperatures, are plotted in Fig.4. At low temperatures both distributions are
almost indistinguishable. Notice the strong sensitivity of the distribution to
temperature variations. When temperatures of order $\Tc$ are reached (cf part
d)), the universality is lost. For $T \sim \Tc$ the fluctuations are dominated
by short orbits, and are system specific. The moments of the distribution can
be computed by the same techniques used in \S \ref{sec:moments} for the grand
potential replacing ${\cal A}_{p,r}$ in Eq.(\ref{apr}) by $ {\cal A}_{p,r} (T)
=\kappa' ( \pi \ T \ r \log p) / (2 r \ p^{r/2} )$.

%%%%%%%%%%%%%%%%%%%%%%%%%%%%%%%%%%%%%%%%%%%%%%%%%%%%%%%%%%%%%%%%%%%

\section{Concluding remarks}
\label{sec:Conclu}
\setcounter{equation}{0}

The use of the Riemannium as a test model in quantum mechanics is justified by
two main reasons. First, by its genericity: the Riemann spectrum possesses all
the generic features of a classically chaotic quantum system with no time
reversal symmetry. Second, by its practical advantages, namely all the
necessary quantum and semiclassical information required to work out
accurate computations and comparisons is available. Hence, though this "number
theoretic" model may seem somewhat remote from a realistic system, it provides
an excellent arena to verify the non trivial quantum mechanical properties of
chaotic systems.

Based on semiclassical techniques and random matrix theory, several aspects of
the thermodynamics of a chaotic Fermi gas have been verified. An accurate
description of the probability distribution of the quantum fluctuations of the
grand potential (or energy) and of the entropy of the Riemannium were
obtained. In particular, the universal linear growth of the entropy variance
followed by a saturation, with a further non universal exponential decay, were
confirmed. The size of the saturation plateau, predicted in the regime
$T_\delta \ll T \ll \Tc$, was relatively small (see the inset in Fig.3). This
is due to the slow logarithmic asymptotic convergence properties of the
Riemannium (in spite of the large chemical potential used in the numerical
simulations, we are not very deep in the semiclassical limit. In fact, for the
window analyzed in the figures the number of particles is around $10^{12}$,
with $g \approx 35$. For an atomic nucleus or for electrons in a metallic
grain, this value of $g$ corresponds to approximately $40\sim 50$ particles.
In the Riemannium, in order to have $T_\delta$ and $\Tc$ separated by a factor
of, say, 100, $\mu$ need to be of the order of $10^{30}$).

The high accuracy of the results obtained for all the quantities studied
confirm the validity of the different approximations employed. The present
theory therefore provides a solid ground to go beyond and test realistic
systems. Of particular interest is the interplay between mean--field
approximations, residual interactions and dynamics. Some encouraging results
in this direction were already obtained in the study of nuclear masses
\cite{mass}.

This work has been supported by the European Commission under the Research
Training Network MAQC (HPRN-CT-2000-00103) of the IHP Programme.

%%%%%%%%%%%%%%%%%%%%%%%%%%%%%%%%%%%%%%%%%%%%%%%%%%%%%%%%%%%%%%%%%%%

\pagebreak

%--------------

\begin{figure}
\begin{center}
\leavevmode
\epsfysize=3.0in
\epsfbox{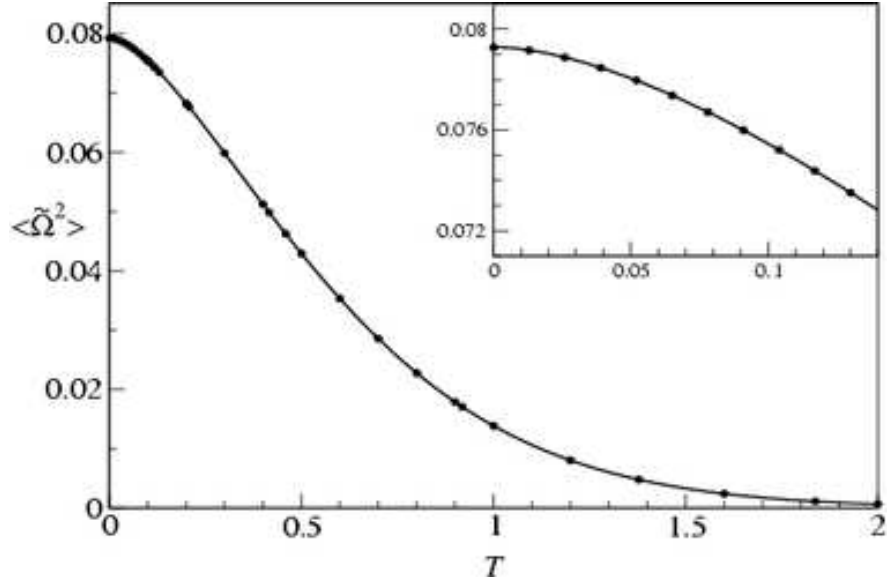}
\vspace{0.5cm}

\caption{{\small Variance of the grand potential fluctuations of the
Riemannium as a function of temperature. Dots: numerical results computed in
the chemical potential window $\mu = (267653402147 \pm 6000)$ containing the
zeros $(10^{12}+1940)$ to $(10^{12}+48684)$. Full line: theoretical prediction
(\ref{varodiag}).}}

\end{center}
\label{VarOmega}
\end{figure}

%-------------

\begin{figure}
\begin{center}
\leavevmode
\epsfysize=5.0in
%\epsfbox{/home/leboeuf/figures/RII_distO.epsf}
\epsfbox{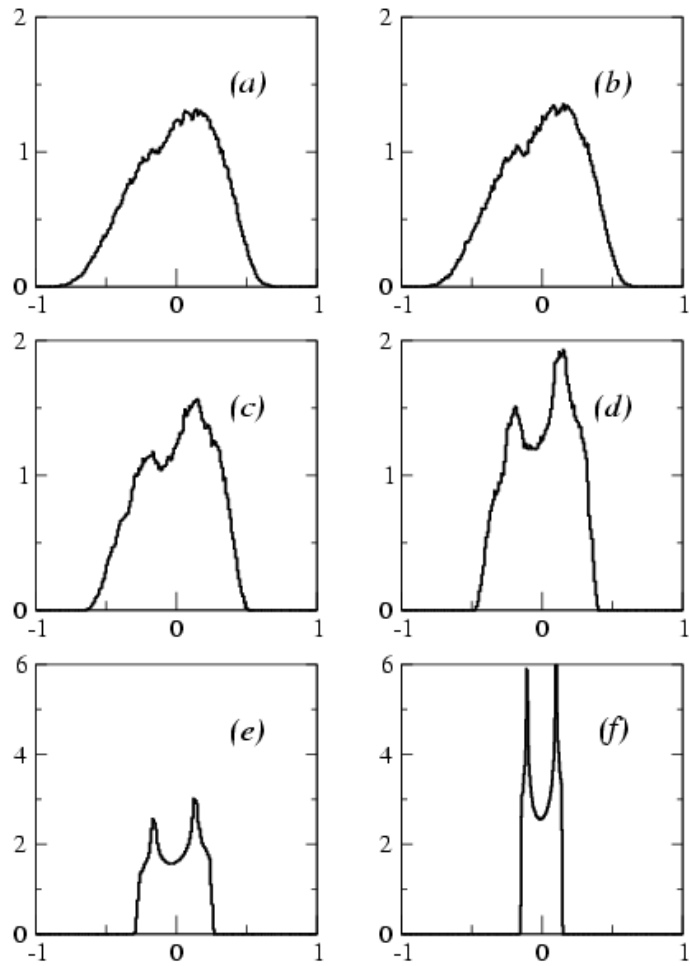}
\vspace{0.5cm}

\caption{{\small Normalized numerical histograms of the probability
distribution of $\widetilde{\Omega}$ at different temperatures: (a) $T = 0$,
(b) $T = 0.1$, (c) $T = 0.3$, (d) $T = 0.5$, (e) $T = 0.8$, (f) $T = 1.2$
(same chemical potential window as in Fig.1).}}

\end{center}
\label{DistOmega}
\end{figure}

%-------------

\begin{figure}
\begin{center}
\leavevmode
\epsfysize=3.0in
%\epsfbox{/home/leboeuf/figures/RII_varS.epsf}
\epsfbox{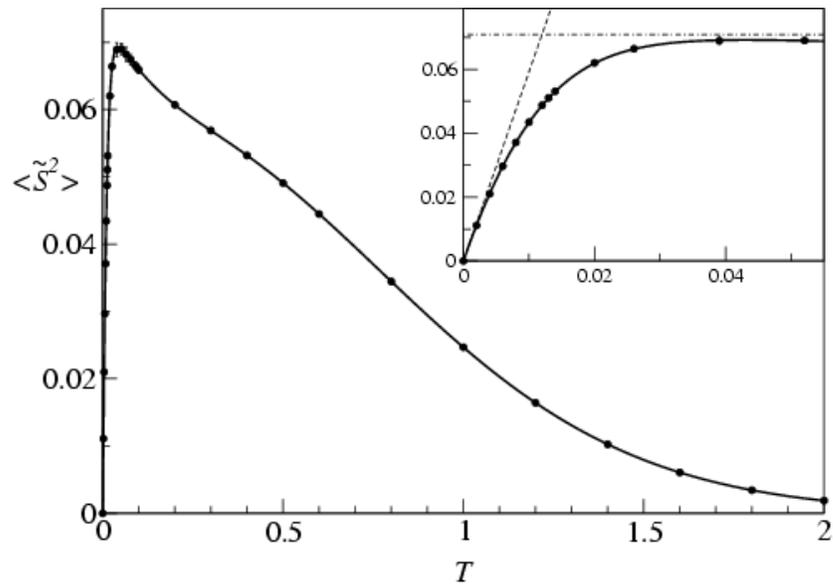}
\vspace{0.5cm}

\caption{{\small Variance of the entropy fluctuations of the Riemannium as a
function of temperature. Dots: numerical results computed in the same chemical
potential window as Fig.1. Full line: theoretical prediction (\ref{S2}).
Dashed line (inset): low temperature approximation (\ref{s2l}). Dot-dashed
line (inset): saturation value (\ref{sat}).}}

\end{center}
\label{VarS}
\end{figure}

%-------------

\begin{figure}
\begin{center}
\leavevmode
\epsfysize=5.0in
%\epsfbox{/home/leboeuf/figures/RII_distS.epsf}
\epsfbox{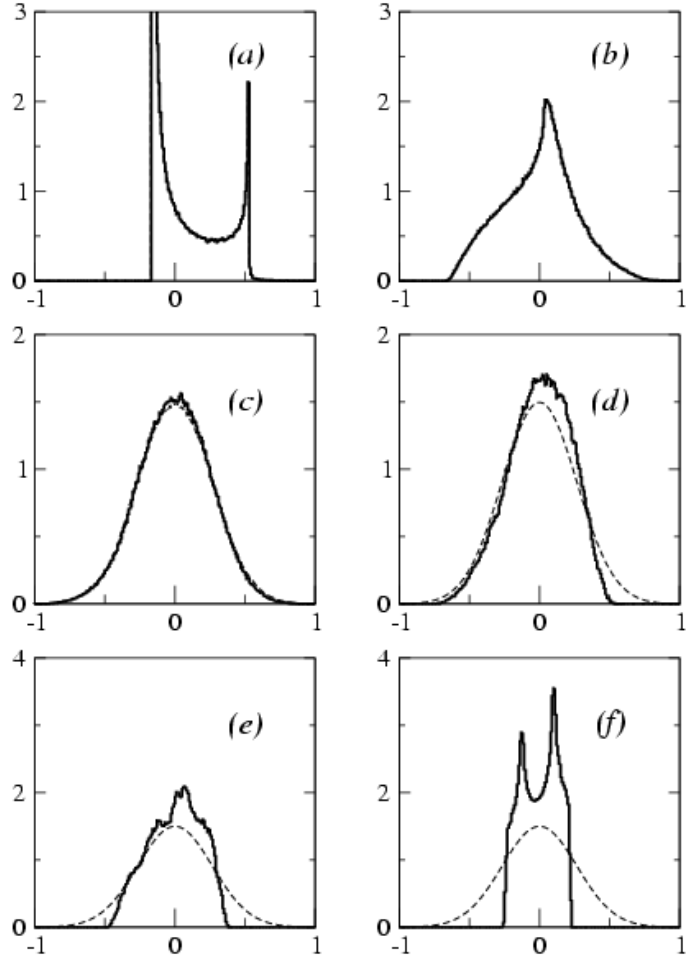}
\vspace{0.5cm}

\caption{{\small Normalized probability distribution of $\widetilde{S}$ at
different temperatures. Full line: numerical results for the Riemannium
computed in the same chemical potential window as Fig.1. Dash line: numerical
results from a GUE single--particle spectrum. (a) $T = T_{\delta} \approx
0.013$, (b) $T = 4 \ T_{\delta} \approx 0.052$, (c) $T = 0.1$, (d) $T = 0.5$,
(e) $T = 0.8$, (f) $T = 1.2$.}}

\end{center}
\label{DistS}
\end{figure}

\

%-------------------

\pagebreak

\begin{center}

\begin{tabular}{|r||r|r|r|}
\hline
~ & T = 0 ~~~~~~~~~~~  & T = 0.3 ~~~~~~~~~~~ & T = 0.5 ~~~~~~~~~~~ \\
\hline
$\langle \widetilde{\Omega}^2 \rangle$ & $(7.928 \pm 0.002) \times
10^{-2}$ & $(5.9885 \pm 0.0025) \times 10^{-2}$ & $(4.2953 \pm 0.0015)
\times 10^{-2}$ \\
~ & $7.9290 \times 10^{-2}$ & $5.9886 \times 10^{-2}$ & $4.2953 \times
10^{-2}$ \\
\hline
$\langle \widetilde{\Omega}^3 \rangle$ & $-(5.78 \pm 0.02) \times
10^{-3}$ & $-(3.44 \pm 0.02) \times 10^{-3}$ & $-(1.65 \pm 0.01)
\times 10^{-3}$ \\
~ & $-5.7822 \times 10^{-3}$ & $-3.4377 \times 10^{-3}$ & $-1.6508
\times 10^{-3}$ \\
\hline
$\langle \widetilde{\Omega}^4 \rangle$ & $(1.480 \pm 0.002) \times
10^{-2}$ & $(7.625 \pm 0.005) \times 10^{-3}$ & $(3.586 \pm 0.003)
\times 10^{-3}$ \\
~ & $1.4814 \times 10^{-2}$ & $7.6273 \times 10^{-3}$ & $3.5869 \times
10^{-3}$ \\
\hline

\end{tabular}

\end{center}

{\small Table 1: Moments of $\widetilde{\Omega}$ at different temperatures.
The upper values (with the errors) were obtained from the numerical
distributions of figure (2.a), (2.c) and (2.d). The lower values are the
semiclassical results (\ref{varodiag}), (\ref{m3}) and (\ref{m4}),
respectively.}\\

%-------------------


\begin{thebibliography}{99}
\bibitem{foot} Three unusual features of the Riemann dynamics are the
following: a) the independence of the periods with energy. This can be
circumvented by interpreting the logarithm of the prime numbers as lengths,
instead of periods; b) contrary to semiclassical approximations, the
oscillating part is here exact (no correction terms are present) and, c) the
negative sign in front of Eq.(\ref{rhoosc}).
\end{thebibliography}
\end{document}